\documentclass[pra,twocolumn]{revtex4-1}

\usepackage{graphicx,bm,amsmath}
\usepackage{color}

\begin{document}

\title{(Spin-)density-functional theory for open-shell systems: exact magnetization density functional for the half-filled Hubbard trimer}

\author{Carsten A. Ullrich}
\affiliation{Department of Physics and Astronomy, University of Missouri, Columbia, Missouri 65211, USA}

\date{\today }

\begin{abstract}
According to the Hohenberg-Kohn theorem of density-functional theory (DFT), all observable quantities of systems of interacting electrons can be expressed as functionals
of the ground-state density. This includes, in principle, the spin polarization (magnetization) of open-shell systems; the explicit form of the
magnetization as a functional of the total density is, however, unknown. In practice, open-shell systems
are always treated with spin-DFT, where the basic variables are the spin densities. Here, the relation between DFT and spin-DFT for open-shell systems
is illustrated and the exact magnetization density functional is obtained for the half-filled Hubbard trimer. Errors arising from spin-restricted
and -unrestricted exact-exchange Kohn-Sham calculations are analyzed and partially cured via the exact magnetization functional.
\end{abstract}

\maketitle

\newcommand{\bfr}{{\bf r}}
\newcommand{\bfm}{{\bf m}}
\newcommand{\bfB}{{\bf B}}
\newcommand{\ua}{\uparrow}
\newcommand{\da}{\downarrow}
\newcommand{\uu}{\uparrow \uparrow}
\newcommand{\ud}{\uparrow \downarrow}
\newcommand{\du}{\downarrow \uparrow}
\newcommand{\dd}{\downarrow \downarrow}
\newcommand{\ME}[1]{\langle \hat #1 \rangle }
\newcommand{\MBZ}{\langle \hat B_z \rangle }
\newcommand{\MBXY}{\langle \hat B_x + \hat B_y \rangle }

\section{Introduction: DFT versus SDFT}

Spin-density-functional theory (SDFT) \cite{Parryang,Dreizlergross,Engel2011}
is concerned with interacting $N$-electron systems in the presence of static magnetic fields. The magnetic fields are assumed to act only on the spins of
the electrons (and not on the orbital currents), giving rise to a Zeeman-like term in the many-body Hamiltonian.
The majority of applications of SDFT are for situations of collinear magnetism; spin then becomes a good quantum number associated with a fixed
spin quantization axis, which is usually chosen to be the $z$-axis. In this case, the basic variables of SDFT are the spin densities,
\begin{equation}\label{spindensity}
n_\sigma(\bfr) = \langle \Psi | \hat \psi^\dagger_\sigma(\bfr) \hat\psi_\sigma(\bfr| \Psi \rangle,
\end{equation}
where $\Psi$ is the many-body wave function, and $\hat\psi^\dagger_\sigma(\bfr)$ and $\hat\psi_\sigma(\bfr)$
are creation and annihilation operators for Fermions of spin $\sigma = \ua,\da$.

The basic theorem of SDFT establishes that in an $N$-electron system under the influence of a scalar potential $V(\bfr)$ and a magnetic field
along the $z$ direction, $B_z(\bfr)$, the
many-body Hamiltonian and all quantities that follow from it are functionals of the ground-state spin densities $n_\sigma(\bfr)$ \cite{Barth1972,Gunnarsson1976,Gidopoulos2007}.
In practice, the spin densities are obtained by solving the following Kohn-Sham equation:
\begin{eqnarray}\label{uKS}
\lefteqn{\hspace{-3.3cm} \left[-\frac{\nabla^2}{2} + V_\sigma(\bfr) + \int \frac{n(\bfr')}{|\bfr - \bfr'|} d\bfr' + V_{{\rm xc},\sigma}[n_\ua,n_\da](\bfr)\right]
\varphi^u_{j\sigma}(\bfr)} \nonumber\\
& =& \varepsilon^u_{j\sigma}\varphi^u_{j\sigma}(\bfr) \:,
\end{eqnarray}
using atomic units with $e=m=\hbar = 1$. Here, the spin-dependent external potential is defined as $V_{\ua,\da}(\bfr) = V(\bfr) \pm \mu_B B_z(\bfr)$,
where $\mu_B$ is the Bohr magneton,
and the exchange-correlation (xc) potential $V_{{\rm xc},\sigma}[n_\ua,n_\da]$ is a functional of the spin densities.

Equation (\ref{uKS}) is known as {\em spin-unrestricted} Kohn-Sham equation \cite{Jacob2012} since, in general, the $\varphi^u_{j\ua}(\bfr)$ and $\varphi^u_{j\da}(\bfr)$ can
be different. From the Kohn-Sham spin orbitals one obtains, in principle, the exact spin densities,
\begin{equation}\label{spinden}
n_\sigma(\bfr) = \sum_{j=1}^{N_\sigma} |\varphi^u_{j\sigma}(\bfr)|^2 \:,
\end{equation}
where $N_\sigma$ is the number of electrons with spin $\sigma$, and $N_\ua + N_\da = N$. The total particle density then follows as
\begin{equation}\label{den}
n(\bfr) = n_\ua(\bfr) + n_\da(\bfr) \:,
\end{equation}
and the magnetization density along the $z$-direction is given by
\begin{equation}\label{mag}
m_z(\bfr) = n_\ua(\bfr) - n_\da(\bfr) \:.
\end{equation}
Notice that the definition $m_z(\bfr) = -\mu_B [n_\ua(\bfr) - n_\da(\bfr)]$ is more frequently used in the literature. Here, however,
we find it convenient, following Engel and Dreizler \cite{Engel2011}, to omit the minus sign and the Bohr magneton $\mu_B$ in the definition of $m_z$.

The SDFT formalism outlined above is suitable for systems of electrons in magnetic fields, but, more often, it is applied to situations
where external magnetic fields are absent and the system is spontaneously magnetic. This happens, for instance, in open-shell atoms and molecules with
an odd number of electrons. In that case, $V_\ua = V_\da = V$, but in general $V_{\rm xc,\ua} \ne V_{\rm xc,\da}$.
Most modern approximate xc functionals are therefore constructed in terms of the spin densities and/or the (unrestricted) spin orbitals \cite{Perdew2003,Cohen2012,Becke2014}.

From a strictly formal point of view, using SDFT in systems without external magnetic field is unnecessary. According to the Hohenberg-Kohn theorem
of DFT \cite{Hohenberg1964,Kohn1965}, the Hamiltonian and all observables in such an $N$-electron systems are functionals of the ground-state
density $n(\bfr)$ alone, which can be obtained from the {\em spin-restricted} Kohn-Sham equation
\begin{eqnarray}\label{rKS}
\lefteqn{\hspace{-3cm}\left[-\frac{\nabla^2}{2} + V(\bfr) + \int \frac{n(\bfr')}{|\bfr - \bfr'|} d\bfr' + V_{{\rm xc}}[n](\bfr)\right]
\varphi^r_{j\sigma}(\bfr)} \nonumber\\
&=& \varepsilon^r_{j\sigma}\varphi^r_{j\sigma}(\bfr) \:,
\end{eqnarray}
where $\varphi^r_{j\ua}(\bfr) = \varphi^r_{j\da}(\bfr)\equiv \varphi_{j}(\bfr)$ and $\varepsilon^r_{j\ua} = \varepsilon^r_{j\da}\equiv \varepsilon_{j}$.

Let us first consider the straightforward case where the system is finite with
an even number of electrons, which is the case for closed-shell configurations. Since then $N_\ua = N_\da = N/2$, the ground-state density is given by a sum over doubly occupied orbitals:
\begin{equation}\label{neven}
n(\bfr) = 2\sum_{j=1}^{N/2} |\varphi_j(\bfr)|^2 \:.
\end{equation}
If, however, the number of electrons $N$ is odd, then the density must be computed as
\begin{equation}\label{nodd}
n(\bfr) = 2\sum_{j=1}^{(N-1)/2} |\varphi_j(\bfr)|^2  + |\varphi_{(N+1)/2}(\bfr)|^2 \:.
\end{equation}
In other words, the first $N-1$ electrons are in doubly occupied orbitals, and the last ($N^{\rm th}$) electron is in a singly occupied orbital.
The spin of individual Kohn-Sham electrons is not explicitly referenced; all electrons experience the same Kohn-Sham effective potential.
If the exact xc potential $V_{\rm xc}[n](\bfr)$
is used in Eq. (\ref{rKS}), then the density $n(\bfr)$ obtained from Eqs. (\ref{neven}) (for even $N$) or (\ref{nodd}) (for odd $N$) will be exact,
and equal to the density obtained via SDFT by using Eq. (\ref{den}).

Within spin-restricted Kohn-Sham, and for odd $N$, the magnetization along $z$ is given by
\begin{equation} \label{magr}
m_z^r(\bfr) = \pm|\varphi_{(N+1)/2}(\bfr)|^2 \:,
\end{equation}
where the sign indicates whether the highest singly occupied orbital is spin-up $(+)$ or spin-down $(-)$.
In general, the restricted Kohn-Sham magnetization is {\em not} equal to the exact magnetization, even if the exact  $V_{\rm xc}[n](\bfr)$
is used in Eq. (\ref{rKS}):
\begin{equation}
m_z^r(\bfr)  \ne m_z(\bfr) \:.
\end{equation}
On the other hand, since in DFT the many-body wave function $\Psi[n]$ is a functional of the density,
the exact magnetization can also be formally expressed as a functional of the density,
$m_z[n](\bfr)$. This means that, in principle, the spin-restricted Kohn-Sham formalism of DFT is sufficient to calculate the magnetization exactly
(again, no external magnetic field is present). However, the form of this density functional $m_z[n](\bfr)$ is unknown.
Therefore, treating open-shell, spin-polarized systems with DFT rather than SDFT is generally discouraged, unless there are specific reasons such as
avoiding the so-called spin contamination problem \cite{Jacob2012,Perdew2009}.

These differences between DFT and SDFT are well known (see also the recent discussion in Ref. \cite{Trushin2018}). Nevertheless, there seems to be no
example in the literature where the magnetization density functional $m_z[n](\bfr)$ is explicitly constructed
and compared with the restricted spin polarization $m_z^r(\bfr)$. In this paper, we will present such a case study,
using the half-filled Hubbard trimer as model system. By comparing exact results with approximate (exchange-only) Kohn-Sham calculations,  errors of
the magnetization will be analyzed and shown to arise from different sources, depending on whether the Kohn-Sham scheme is restricted or unrestricted.
The exact magnetization functional will be shown to provide a cure to some of the errors, but not all of them.

\begin{figure*}[t]
\includegraphics[width=\linewidth]{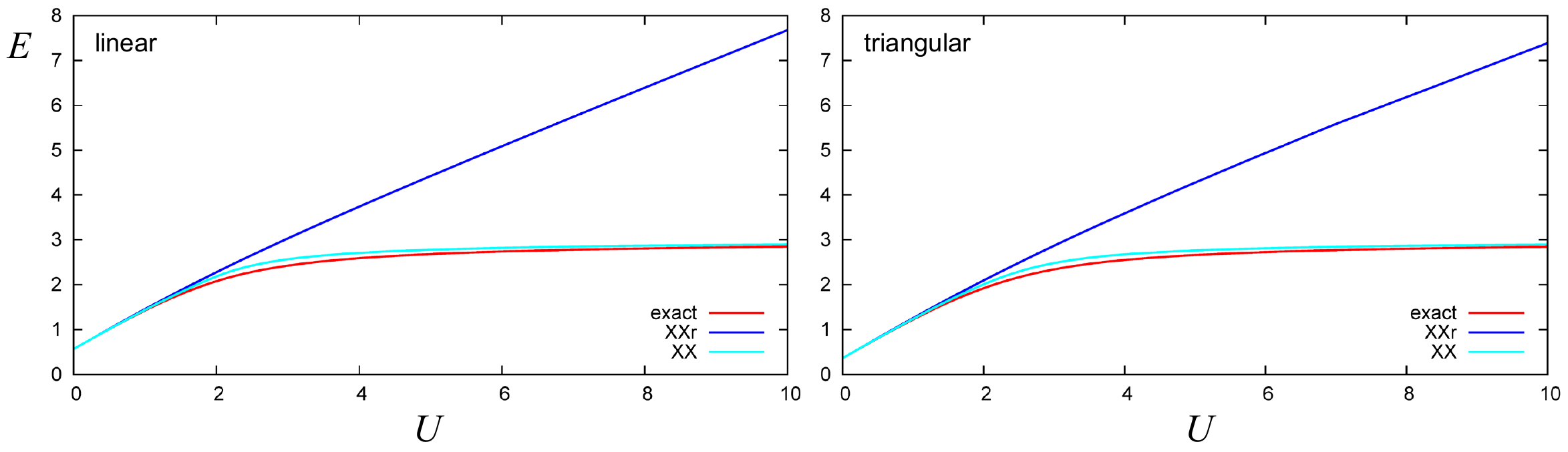}
\caption{(Color online)
Total ground-state energies of the linear and triangular asymmetric half-filled Hubbard trimer with $V_1=1$, $V_2=0$, and
$V_3=2$, comparing exact results with XX and XXr Kohn-Sham calculations.} \label{fig1}
\end{figure*}

\section{The half-filled Hubbard trimer}

\subsection{Model}
The inhomogeneous Hubbard model is defined by the following Hamiltonian \cite{Capelle2013}:
\begin{eqnarray}\label{Hubbard}
\hat H &=& -t\sum_{\langle i,j \rangle} \sum_{\sigma} (\hat c^\dagger_{i\sigma}\hat c_{j\sigma} + \hat c^\dagger_{j\sigma}\hat c_{i\sigma}) + U\sum_{j}
\hat c^\dagger_{j\ua}\hat c_{j\ua}\hat c^\dagger_{j\da}\hat c_{j\da} \nonumber\\
&+& \sum_{j}  V_j \sum_{\sigma} \hat c^\dagger_{j\sigma}\hat c_{j\sigma} \:,
\end{eqnarray}
where $\hat c_{j\sigma}^\dagger$ and $\hat c_{j\sigma}$ are creation and annihilation operators, respectively,
for electrons with spin $\sigma$ on site $j$, and $\langle i,j\rangle$ indicates pairs of nearest-neighbor lattice sites.
We fix the hopping parameter as $t=0.5$. The on-site interaction strength $U$
and the external potential $V_j$ will be treated as variable parameters.

The corresponding Kohn-Sham Hamiltonian is
\begin{equation}\label{HubbardKS}
\hat H^{\rm KS} = -t\sum_{\langle i,j \rangle} \sum_{\sigma} (\hat c^\dagger_{i\sigma}\hat c_{j\sigma} + \hat c^\dagger_{j\sigma}\hat c_{i\sigma})
+  \sum_{j} \sum_{\sigma} \hat V_{j\sigma}^{\rm KS} c^\dagger_{j\sigma}\hat c_{j\sigma} \:,
\end{equation}
where the Kohn-Sham potential at lattice site $j$ is given by $V_{j\sigma}^{\rm KS} = V_j+V_j^{\rm H}+V_{j\sigma}^{\rm xc}$.
The Hartree potential is simply
\begin{equation} \label{Hartree}
V_j^{\rm H} = U n_j \:,
\end{equation}
where $n_j$ is the density (site occupation) on the $j^{\rm th}$ lattice point. In the spin-restricted Kohn-Sham scheme, the xc potential is
constrained to satisfy $V_{j\ua}^{\rm xc} = V_{j\da}^{\rm xc}$.

For the Hubbard model with on-site interactions, the simplest approximation for the xc potential is exact exchange (XX), given by \cite{Ullrich2018}
\begin{equation}\label{XX}
V_{j\sigma}^{\rm XX} = -U n_{j\sigma} \:.
\end{equation}
The spin-restricted exact exchange (XXr) potential is
\begin{equation}\label{XXr}
V_{j}^{\rm XXr} = -U n_{j}/2\:.
\end{equation}
To include correlation effects, one could use the Bethe-ansatz local density approximation (BALDA) \cite{Capelle2013,Lima2002,Lima2003,Franca2012}. However,
the BALDA runs into convergence problems on small lattices whenever any of the site occupations approaches 1 \cite{Ullrich2018}.
Alternatively, correlation can be treated via orbital functionals, as discussed elsewhere \cite{Ullrich2018,Pluhar2019}.
In this paper, however, correlation effects are not included to keep things simple.

The Hubbard model, especially the half-filled Hubbard dimer, has been widely used as a test system for DFT \cite{Carrascal2015,Cohen2016,Carrascal2018,Franca2018,Herrera2018,Ullrich2018}.
In the absence of an external magnetic field, the ground state of the Hubbard dimer is a singlet, with zero total spin and no magnetization.
For the purpose of comparing DFT and SDFT, the simplest suitable system is the half-filled Hubbard trimer, that is, three electrons on a three-point lattice
\cite{footnote}.
The half-filled Hubbard trimer has been discussed in the literature \cite{Shiba1972,Trif2010,Nossa2012,Tabrizi2019}, but not in the
context of DFT. Most notably, earlier studies of the Hubbard trimer were limited to the homogeneous case of constant potential. Here, by contrast,
we consider the inhomogeneous case.

\subsection{Example: from weakly to strongly correlated} \label{sec:IIB}

\begin{figure}[t]
\includegraphics[width=\linewidth]{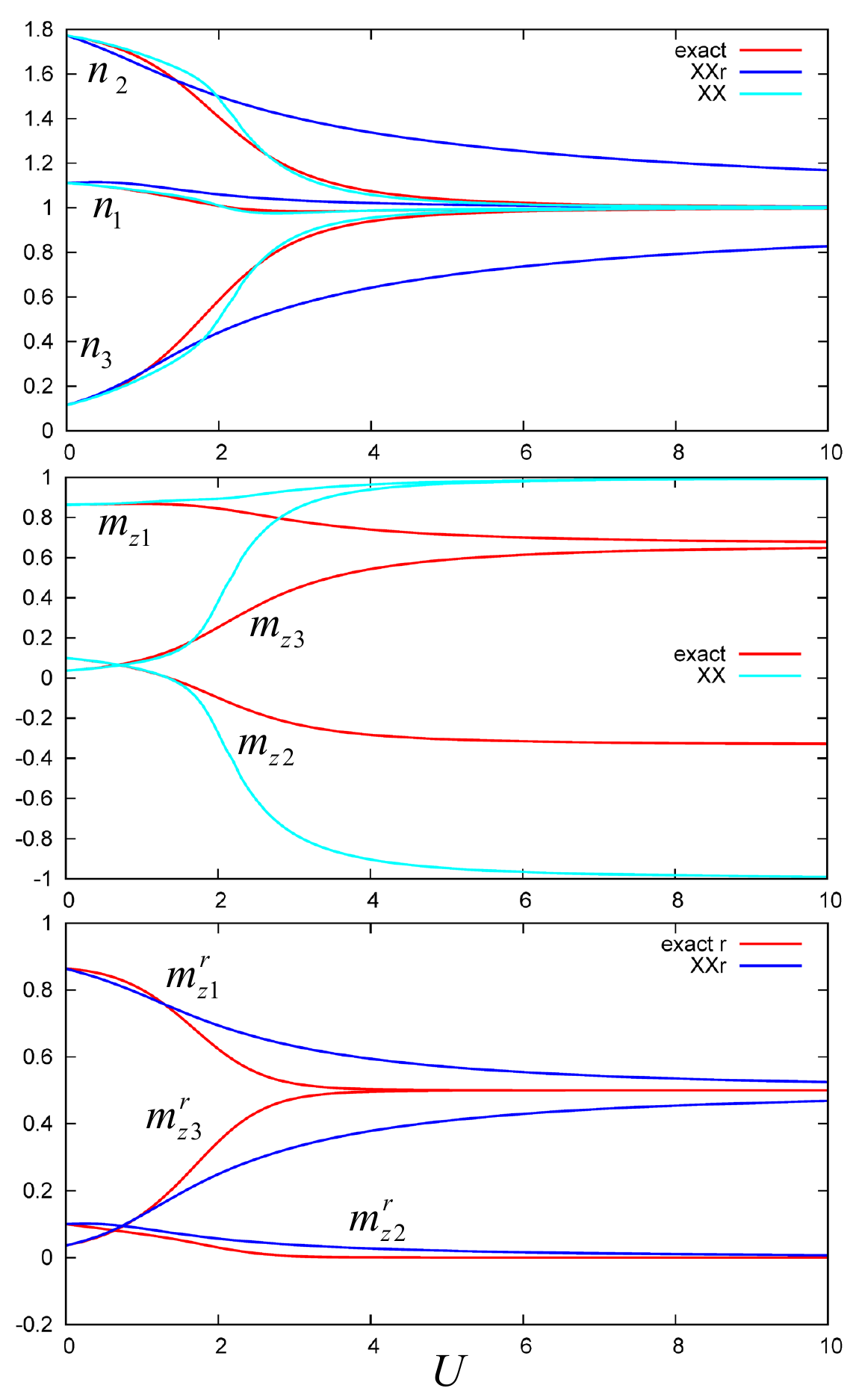}
\caption{(Color online)
Density $n_j$ (top panel) and magnetization $m_{zj}$ (middle and bottom panels), on lattice points $j=1,2,3$ of a linear half-filled Hubbard trimer with $V_1=1$, $V_2=0$, and
$V_3=1$. Exact results are compared with spin-restricted and -unrestricted Kohn-Sham results using XX (see text for further details).} \label{fig2}
\end{figure}

\begin{figure}[t]
\includegraphics[width=\linewidth]{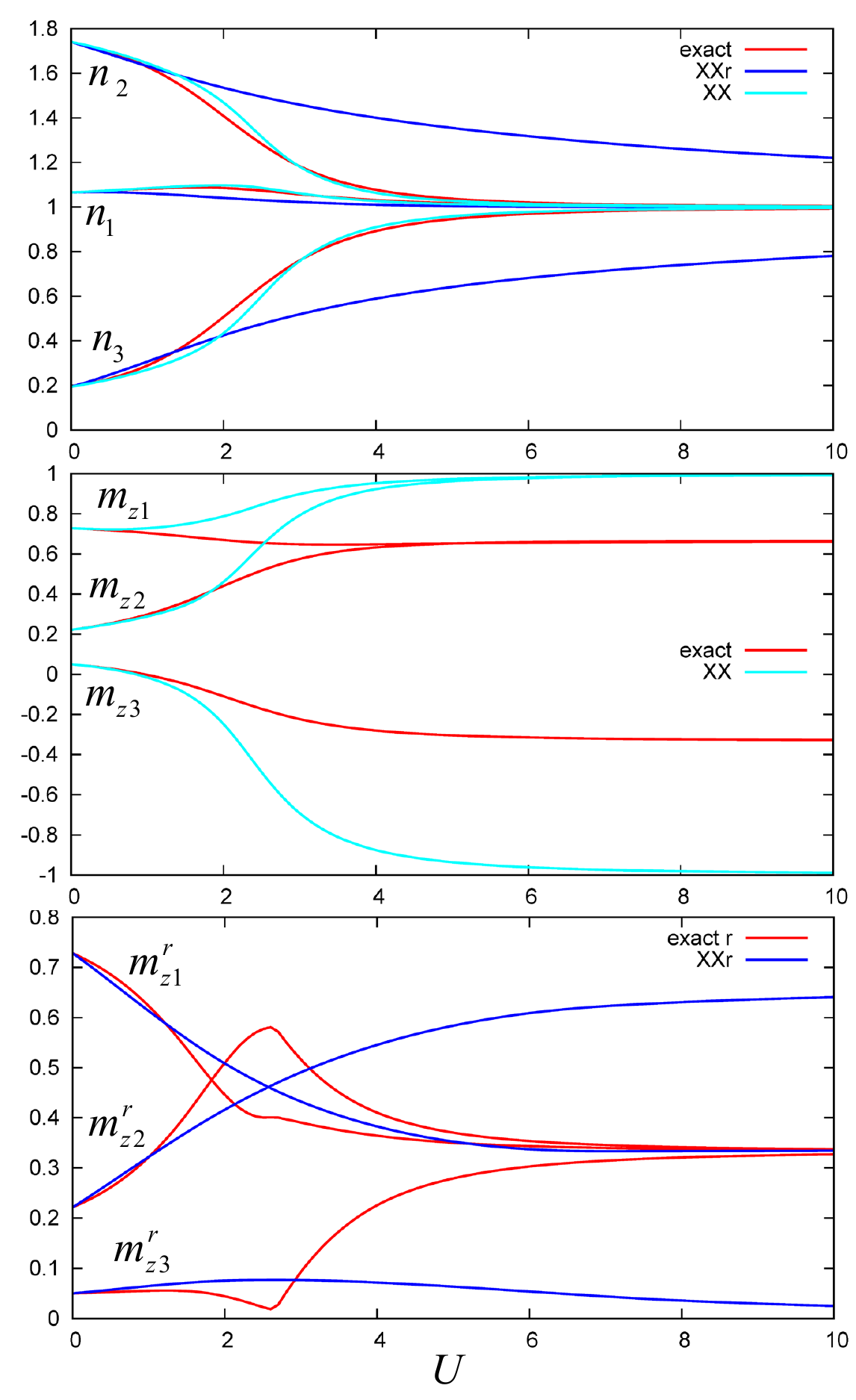}
\caption{(Color online)
Same as Fig. \ref{fig2}, but for a triangular lattice.} \label{fig3}
\end{figure}

\subsubsection{Ground-state energies and densities}

To illustrate the half-filled Hubbard trimer in the weakly and strongly correlated limits, we now consider an asymmetric example where we
fix the external potential as $V_1=1$, $V_2 = 0$,  and $V_3=2$ and vary the interaction strength from $U=0$ to 10.
We obtain the ground state of the interacting system (in both linear and triangular configuration) through exact diagonalization; technical details
are given in the Appendix. This yields the exact ground-state energy $E$, density $n_j$ and magnetization $m_{zj}$ for each given $U$.

We compare the exact solutions with restricted and unrestricted Kohn-Sham calculations using XX and XXr, respectively.
Figure \ref{fig1} shows the ground-state energies. The results are very similar for the linear and the triangular lattice:
in both cases, XX agrees closely with the exact ground-state energy, approaching a final limit of $E=3$ as the electrons localize after a clear crossover
from weakly to strongly correlated around $U=2$.  The spin-restricted XXr Kohn-Sham calculation, on the other hand, produces a continuously increasing energy.
This familiar behavior was observed earlier for the case of the half-filled (spin singlet) Hubbard dimer \cite{Carrascal2015}. There, however,
the spin symmetry had to be artificially broken to get better agreement with the exact results.

The ground-state densities are shown in the top panels of Fig. \ref{fig2} for the linear trimer and of Fig. \ref{fig3} for the triangular lattice.
For XX and XXr, the density follows from Eqs. (\ref{den}) and (\ref{nodd}), respectively.
The exact solution shows that the density behaves very similarly in the two cases: for small $U$,
the density is much larger on the second lattice point (where the potential is $V_2=0$) compared to the two other lattice points (where $V_1=1$ and $V_3=2$).
As $U$ increases, the electrons begin to localize; eventually, for $U\sim 10$, all lattice points are equally populated ($n_1=n_2=n_3=1$).
The crossover between the delocalized and localized regimes is not very sharp, but occurs in a region roughly between $U=1$ and $3$.

\begin{figure*}[t]
\includegraphics[width=\linewidth]{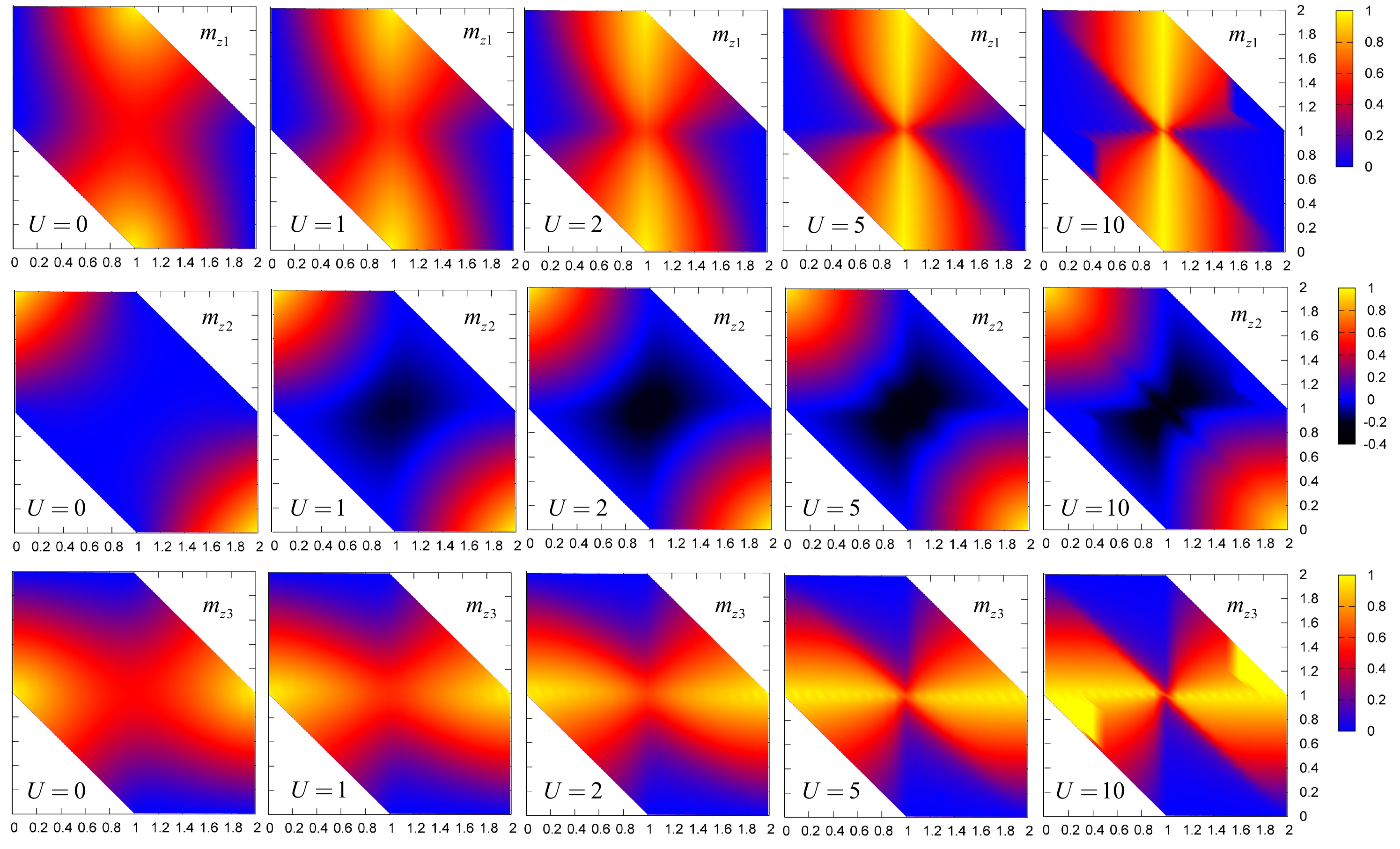}
\caption{(Color online)
Magnetization functional $m_z[n]$ for the linear half-filled Hubbard trimer, and for different strengths of the interaction $U$.
The top, middle and bottom panels show $m_{zj}$ for the lattice points $j=1,2,3$. In the color maps, the horizontal axis is $n_1$ and the vertical axis is $n_3$.
} \label{fig4}
\end{figure*}

\begin{figure*}[t]
\includegraphics[width=\linewidth]{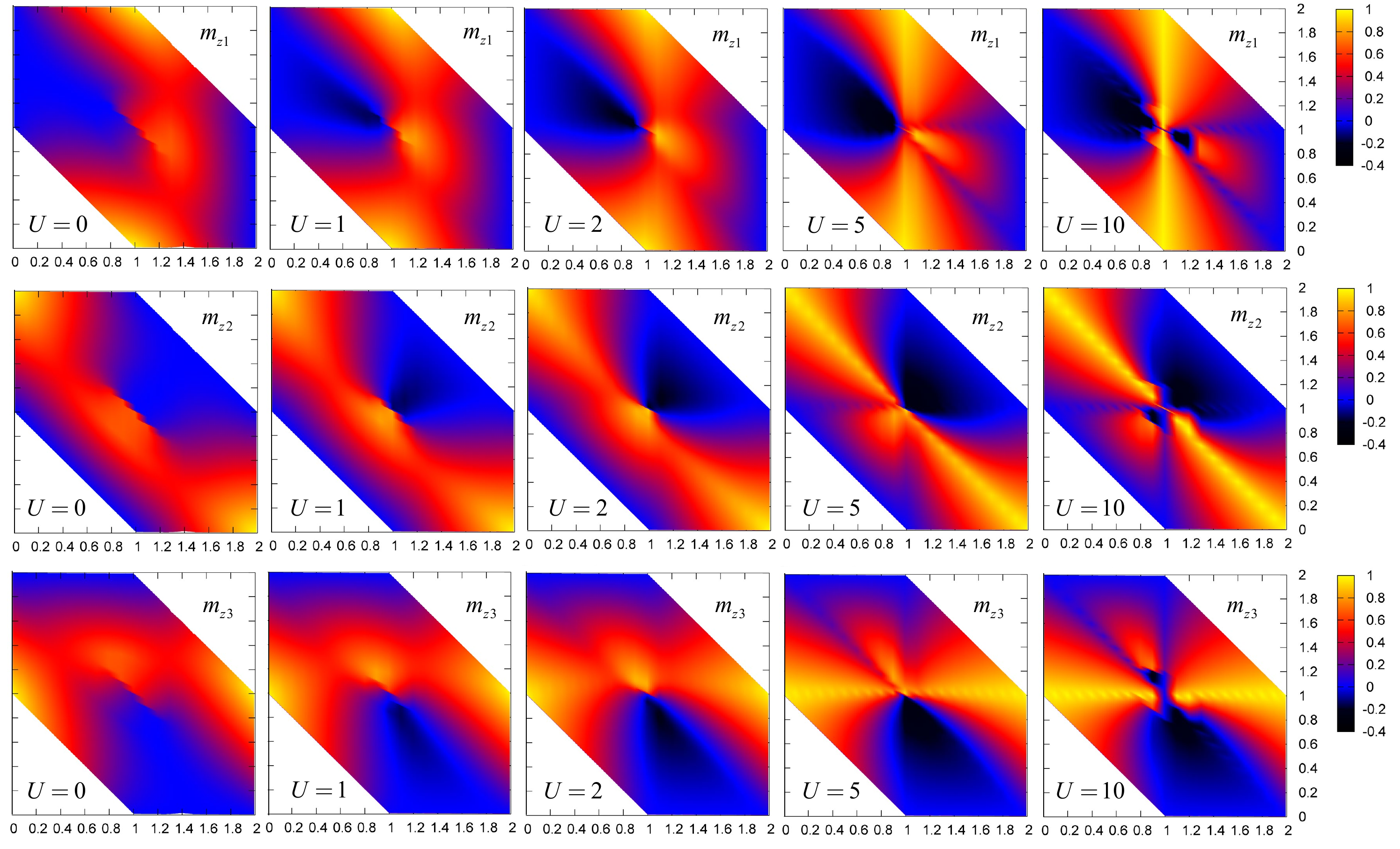}
\caption{(Color online)
Same as Fig. \ref{fig4}, but for a triangular lattice.} \label{fig5}
\end{figure*}

This behavior of the density is very well reproduced by the spin-unrestricted XX calculations. On the other hand, XXr
agrees less well with the exact density: on both lattices, the transition from delocalized to localized (where all $n_j\to 1$) is
too slow.

\subsubsection{Magnetizations}

The results for the magnetization are shown in the middle and bottom panels of Figs. \ref{fig2} and \ref{fig3}.
For XX, the magnetization is obtained from the approximate spin densities without further approximation, using Eq. (\ref{mag}); in the case of XXr,
the magnetization is obtained using Eq. (\ref{magr}), which is an approximation, even if the density and the (restricted) Kohn-Sham orbitals were exact.

The middle panels of Figs. \ref{fig2} and \ref{fig3} show the exact $m_z$, compared with unrestricted XX. Again, the linear and triangular lattices behave in a
similar manner. The exact magnetization is such that, for small $U$, the first lattice point is almost fully magnetized, whereas the second and third lattice
point have almost no magnetization. As $U$ increases, both lattices develop a slightly antiferromagnetic pattern: for the linear lattice,
$m_{z1}$ and $m_{z3}$ approach $2/3$ and $m_{z2}$ approaches $-1/3$; for the triangular lattice the roles of $m_{z2}$ and $m_{z3}$ are reversed.
It is interesting to observe that the localization of the density implies equal site occupation, but not equal magnetization, even in the case
of the triangular lattice with periodic boundary conditions.

This behavior is reasonably well reproduced by the spin-unrestricted XX calculations. The agreement is excellent for both lattices up until $U \sim 2$, but
for stronger interactions the antiferromagnetism is drastically overestimated: on two lattice points the magnetization approaches 1 and on the third point it
approaches $-1$. This demonstrates that an SDFT calculation can produce excellent total densities and energies, but poor spin densities.
This well-known phenomenon has been observed, for example, in transition-metal complexes \cite{Jacob2012,Conradie2007,Boguslawski2011,Song2018}.

The bottom panels of Figs. \ref{fig2} and \ref{fig3} compare the exact restricted magnetization $m_z^r$ with XXr.
The exact $m_z^r$ is obtained by first calculating the exact density for a given $U$, then determining that potential which, for $U=0$, reproduces
this density (this is the exact restricted $V_j^{\rm KS}$), and then calculating the restricted magnetization from this (see Section \ref{sec:IIC} for further details
of the inversion procedure).

Comparing $m_z^r$ with the exact $m_z$ reveals significant differences for moderate and high interaction strengths:
mainly, the antiferromagnetic trend is not reproduced. For the linear lattice, $m_{z1}^r$ and $m_{z3}^r$ approach $1/2$ instead of $2/3$, and $m_{z2}^r$
goes to zero instead of $-1/3$. The XXr calculations agree quite well with the exact $m_z^r$ for all interaction strengths.

On the triangular lattice, the exact $m_z^r$ approaches a completely symmetric strongly-correlated limit at large $U$, with $m_{zj}^r\to 1/3$ for all lattice points.
The crossover region exhibits quite dramatic changes of the behavior of the magnetization, which has an interesting underlying cause:
from $U=2.7$ onwards, the second and third Kohn-Sham level become degenerate, which requires determining the correct linear combination
of the degenerate orbitals to reproduce the exact density via Eq. (\ref{nodd}). The associated magnetization, Eq. (\ref{magr}), is strongly affected by this
choice of orbital. A similar degeneracy occurs in the XXr calculations,
but only around $U=7$, and with no dramatic consequences for the magnetization.
Thus, the agreement between XXr and exact $m_z^r$ is good for small $U$, but XXr approaches an unsymmetric limit at large $U$, with
$m_{z1}\to 1/3$, $m_{z2}\to 2/3$, and $m_{z3}\to 0$;  incidentally, this happens to be closer to the exact unrestricted $m_z$ than to the exact $m_z^r$.

The examples discussed here are typical for the behavior of the inhomogeneous half-filled Hubbard trimer. In general, XX, both in its spin-restricted and
-unrestricted form, performs well as long as the system is weakly correlated, but fails for the magnetization at strong correlation.
However, the failures of the restricted and unrestricted
XX are of a qualitatively different nature, providing prime examples of the two different kinds of errors of (S)DFT, namely, those caused by approximations
of the functionals, and those caused by using approximate densities (``density-driven errors'') \cite{Song2018,Kim2013,Sim2018}. In the following, we will
investigate these errors more closely, looking for ways in which they could possibly be mitigated.

\subsection{The magnetization density functional} \label{sec:IIC}

We now construct the functional $m_z[n]$ which produces the exact magnetization from a given ground-state density, for the half-filled Hubbard trimer with
interaction $U$. To do this, we invert the interacting 3-body Schr\"odinger equation on the 3-point lattice (see Appendix), governed by the Hamiltonian (\ref{Hubbard}).
In other words, we construct the lattice potential $V_j$ that produces a given ground-state density distribution $n_j$ (for a given $U$), and from
the solution of the Schr\"odinger equation with this potential we obtain the exact magnetization $m_{zj}$. This is repeated for all possible density distributions,
which then defines the functional $m_z[n]$.

In practice, to determine the potential $V_j$ from the density, we minimize the mean square deviation of
the associated ground-state density $n_j$  from a given target density $n^t_j$,  using Powell's conjugate direction method \cite{Press2007}.
This only requires a moderate computational effort for the systems under study.

The results are shown in Fig. \ref{fig4} for the linear Hubbard trimer and in Fig. \ref{fig5} for the triangle.
In the graphic representation of $m_{zj}[n]$, $n_1$ and $n_3$ are chosen as the two independent values of the density;
$n_2$ follows as $n_2=3-n_1-n_3$, and all densities (or site occupations) are constrained to satisfy $0\le n_j \le 2$.
The domain of the functional is therefore a square area in $n_1-n_3$ space with two triangular regions in the upper right-hand and lower left-hand corners excluded.

Figures \ref{fig4} and \ref{fig5} illustrate the functional $m_{zj}[n]$ on the three lattice points $j=1,2,3$, and for five different interaction
strengths, $U=0,1,2,5,10$. The color maps show that the range of $m_{zj}$ on the Hubbard lattices is between $-1/3$ and 1. The condition
$m_{z1}+m_{z2}+m_{z3}=1$ is always satisfied.
However, the detailed behavior of the magnetization is quite different for the linear chain (fixed boundary conditions) and the triangle (periodic boundary conditions).

For the noninteracting case, the magnetization lies between 0 and 1 on both lattices. As the interaction strength $U$ increases, the maps generally develop sharper features.
In particular, pockets of negative magnetization begin to appear, marked by dark blue turning into black. The partial antiferromagnetism is thus a generic feature of the half-filled Hubbard trimer.
For the linear lattice, only the magnetization on the central site can become negative; for the triangular lattice, it can be negative on either of the three sites.

\subsection{Eliminating the functional error using the exact magnetization density functional}

In Section \ref{sec:IIB}, we saw that the approximate Kohn-Sham calculations are plagued by different kinds of errors.
XX generally leads to good total energies and densities, but the individual spin densities [which give rise to the magnetization via Eq. (\ref{mag})]
are not accurate for moderate and strong correlations. This is a clear example of a density-driven (or, more precisely, spin-density-driven) error,
since the magnetization functional is exact in the unrestricted case.
XXr, on the other hand, is not very accurate for total densities (unless $U$ is small), and
produces magnetizations of poor quality. Hence, the errors are both density-driven and due to a poor approximation of the magnetization functional,
but it is not clear a priori which is worse.

\begin{figure}
\includegraphics[width=\linewidth]{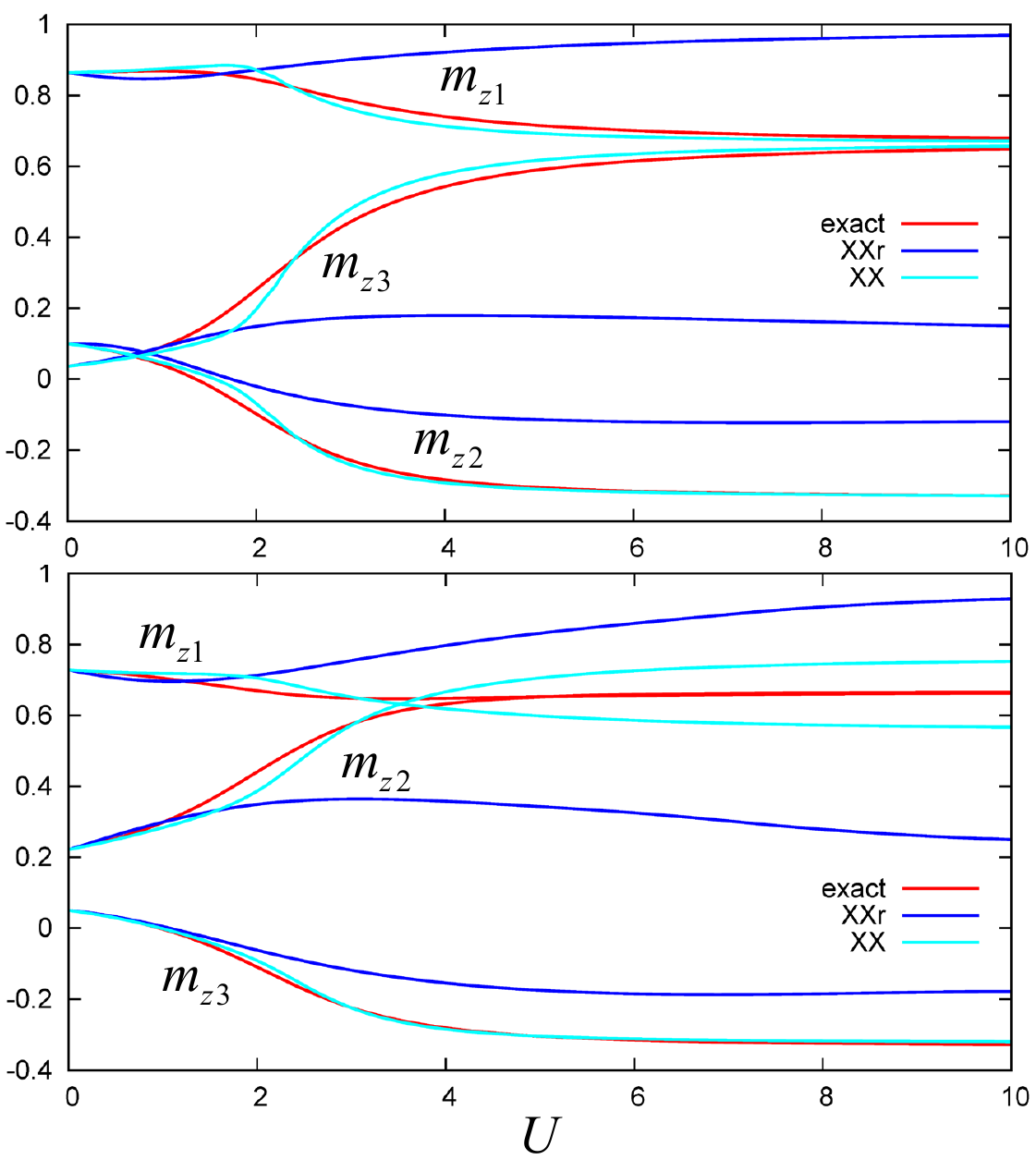}
\caption{(Color online) Magnetization of the linear (top panel) and triangular (bottom panel) lattice considered in Figs. \ref{fig2} and \ref{fig3}, respectively,
obtained by evaluating the exact magnetization functional s $m_{zj}[n]$ with the self-consistent XX and XXr total densities.} \label{fig6}
\end{figure}

Let us now eliminate the error due to the wrong magnetization functional, and focus on the density-driven errors by themselves. Figure \ref{fig6} shows the
magnetizations of the linear and triangular lattices (for the same systems considered in Figs. \ref{fig2} and \ref{fig3}) which follow
from evaluating the exact functional $m_{zj}[n]$ with the self-consistent XX and XXr densities.
The spin-unrestricted XX now agrees very well with the exact magnetization for all interaction strengths, reproducing the weakly antiferromagnetic behavior in both cases.
The spin-density-driven error is thus clearly the dominant one in XX.

On the other hand, XXr, in spite of some improvement, is still not very close to the exact result, particularly for interaction strengths greater than $U=2$.
Overall, the exact magnetization functional cannot make up for the deficiencies of the XXr total density. For better agreement, correlation effects must be 
included on top of XXr.

\section{Conclusions}

In the standard SDFT approach to open-shell systems, the magnetization (or, equivalently, the spin polarization) is expressed
in terms of the spin densities; the latter are obtained using a spin-unrestricted Kohn-Sham formalism, which uses the spin densities as
two independent variables. However, according to the Hohenberg-Kohn theorem, all observable quantities
follow in principle from the total density; an explicitly spin-polarized Kohn-Sham formalism, with two types of orbitals (spin-up and spin-down)
is therefore, in principle, unnecessary. This could reduce the computational effort of such systems by half.
Of course, the reason that SDFT is almost always preferred over DFT is that it is much easier to construct accurate
approximations thanks to the flexibility that is afforded by having two independent variables. Nevertheless, the point of principle remains.

Here, we have constructed and analyzed the exact density functional of the magnetization, $m_z[n]$, for two simple model systems, the half-filled
Hubbard trimer with fixed and periodic boundary conditions (chain and triangle). The functional is rather smooth for small interactions strengths $U$,
but acquires more sharp features in the strongly correlated limit, where the system tends towards a partial antiferromagnetism.

Two approximations were tested: spin-unrestricted and spin-restricted exact exchange (XX and XXr). As expected, XX performs much better, giving very good
total energies and densities, but fails to reproduce the correct magnetization for moderate and strong interactions. The error thus clearly lies in the individual spin densities,
as demonstrated by plugging the XX total density into the exact magnetization functional. On the other hand, XXr fails as soon as the system is no longer
weakly interacting, which leads to poor energies, densities and (restricted) magnetization. This is not surprising, based on what is known from other
test cases such as the half-filled Hubbard dimer. The exact magnetization functional is unable to provide a cure for the shortcomings of XXr.

The case study discussed in this paper thus provides a potentially valuable proof of principle for (S)DFT. Many of today's popular xc
functionals are not optimized to produce the best spin densities, but focus more on total energies and total densities. This can be problematic if
explicitly spin-dependent observables are of interest, such as electron paramagnetic resonance parameters \cite{Boguslawski2011}.
The construction of explicit (but approximate) magnetization functionals depending on total densities may therefore be a worthwhile effort.
The example of the half-filled Hubbard trimer has shown that this can lead to quite significant improvements.

\acknowledgments{This work was supported by Department of Energy Grant No. DE-SC0019109 and by a Research Corporation
for Science Advancement Cottrell Scholar SEED Award.
}

\appendix

\section{Exact diagonalization of the half-filled Hubbard trimer in a magnetic field}
This Appendix gives technical details on the numerical solution of the half-filled Hubbard trimer in the presence of
arbitrary scalar potentials and magnetic fields (acting as Zeeman terms on the spins only).

Consider three interacting electrons on a three-point lattice, with sites 1,2,3, arranged along a line (non-periodic boundary conditions) or as a triangle (periodic boundary
conditions). The scalar potential on the three lattice
sites is denoted by $V_{1,2,3}$, and the magnetic field is denoted by $\bfB_{1,2,3}$.
The general three-particle Schr\"odinger equation reads:
\begin{eqnarray} \label{3electronSE}
\lefteqn{\hspace{-4.8cm}
\left\{ \sum_{i=1}^3 [T(i) + V(i)+ \bfB(i)\cdot {\bm\sigma}(i)] + \sum_{i< j}^3 W(i,j)\right\} \Psi_j(1,2,3)} \nonumber\\
= \hat H(1,2,3) \Psi_j(1,2,3) & &= E_j \Psi_j(1,2,3) \;,
\end{eqnarray}
where $T(i)$ is the kinetic energy operator acting on the $i$th electron (here, nearest-neighbor hopping with amplitude $t$), $\bm \sigma$ is the
vector of Pauli matrices, and $W(i,j)$ is the interaction between the $i$th and $j$th electron (here, on-site interaction with strength $U$).

The wave functions $\Psi_j(1,2,3)$ can be expanded in a basis of Slater determinants, $\{\psi_n(1,2,3)\}$:
\begin{equation} \label{wavefct}
\Psi_j(1,2,3) = \sum_n c_{nj} \psi_n(1,2,3) \;.
\end{equation}
The expansion coefficients $c_{nj}$ and the energies $E_j$ follow from diagonalization
of the Hamiltonian matrix $\bf H$:
\begin{equation}\label{M}
{\bf H} \vec{c}_j = E_j \vec{c}_j \;,
\end{equation}
where the elements of the $20\times 20$ matrix $\bf H$ are given by $ \ME H_{m,n} = \langle \psi_m |\hat H | \psi_n \rangle$ [$\hat H$ is the full three-electron 
Hamiltonian of Eq. (\ref{3electronSE})]. The individual parts of the Hamiltonian matrix $\ME H_{m,n} = \ME T_{m,n} + \ME V_{m,n} + \ME B_{m,n}+ \ME W_{m,n}$ will now be derived.

\begin{figure*}

\noindent\unitlength1cm
\begin{picture}(1,1)
\put(0,0){\circle*{0.15}}
\put(0.6,0){\circle*{0.15}}
\put(1.2,0){\circle*{0.15}}
\put(-0.075,0.3){$\ua$}
\put(0.525,0.3){$\da$}
\put(1.125,0.3){$\da$}
\end{picture}
\hspace{18mm}
\begin{picture}(1,1)
\put(0,0){\circle*{0.15}}
\put(0.6,0){\circle*{0.15}}
\put(1.2,0){\circle*{0.15}}
\put(-0.075,0.3){$\da$}
\put(0.525,0.3){$\ua$}
\put(1.125,0.3){$\da$}
\end{picture}
\hspace{18mm}
\begin{picture}(1,1)
\put(0,0){\circle*{0.15}}
\put(0.6,0){\circle*{0.15}}
\put(1.2,0){\circle*{0.15}}
\put(-0.075,0.3){$\da$}
\put(0.525,0.3){$\da$}
\put(1.125,0.3){$\ua$}
\end{picture}
\hspace{18mm}
\begin{picture}(1,1)
\put(0,0){\circle*{0.15}}
\put(0.6,0){\circle*{0.15}}
\put(1.2,0){\circle*{0.15}}
\put(-0.075,0.3){$\da$}
\put(0.525,0.3){$\ua$}
\put(1.125,0.3){$\ua$}
\end{picture}
\hspace{18mm}
\begin{picture}(1,1)
\put(0,0){\circle*{0.15}}
\put(0.6,0){\circle*{0.15}}
\put(1.2,0){\circle*{0.15}}
\put(-0.075,0.3){$\ua$}
\put(0.525,0.3){$\da$}
\put(1.125,0.3){$\ua$}
\end{picture}
\hspace{18mm}
\begin{picture}(1,1)
\put(0,0){\circle*{0.15}}
\put(0.6,0){\circle*{0.15}}
\put(1.2,0){\circle*{0.15}}
\put(-0.075,0.3){$\ua$}
\put(0.525,0.3){$\ua$}
\put(1.125,0.3){$\da$}
\end{picture}

\noindent
\hspace{4.5mm}$\psi_1$\hspace{26.5mm}$\psi_2$\hspace{26.5mm}$\psi_3$\hspace{26.5mm}$\psi_4$\hspace{26.5mm}$\psi_5$\hspace{26.5mm}$\psi_6$

\vspace{1mm}

\noindent\unitlength1cm
\begin{picture}(1,1)
\put(0,0){\circle*{0.15}}
\put(0.6,0){\circle*{0.15}}
\put(1.2,0){\circle*{0.15}}
\put(-0.15,0.3){$\ud$}
\put(0.525,0.3){$\da$}
\put(1.125,0.3){}
\end{picture}
\hspace{18mm}
\begin{picture}(1,1)
\put(0,0){\circle*{0.15}}
\put(0.6,0){\circle*{0.15}}
\put(1.2,0){\circle*{0.15}}
\put(-0.15,0.3){$\ud$}
\put(0.525,0.3){}
\put(1.125,0.3){$\da$}
\end{picture}
\hspace{18mm}
\begin{picture}(1,1)
\put(0,0){\circle*{0.15}}
\put(0.6,0){\circle*{0.15}}
\put(1.2,0){\circle*{0.15}}
\put(-0.075,0.3){$\da$}
\put(0.45,0.3){$\ud$}
\put(1.125,0.3){}
\end{picture}
\hspace{18mm}
\begin{picture}(1,1)
\put(0,0){\circle*{0.15}}
\put(0.6,0){\circle*{0.15}}
\put(1.2,0){\circle*{0.15}}
\put(-0.075,0.3){}
\put(0.45,0.3){$\ud$}
\put(1.125,0.3){$\da$}
\end{picture}
\hspace{18mm}
\begin{picture}(1,1)
\put(0,0){\circle*{0.15}}
\put(0.6,0){\circle*{0.15}}
\put(1.2,0){\circle*{0.15}}
\put(-0.075,0.3){$\da$}
\put(0.525,0.3){}
\put(1.,0.3){$\ud$}
\end{picture}
\hspace{18mm}
\begin{picture}(1,1)
\put(0,0){\circle*{0.15}}
\put(0.6,0){\circle*{0.15}}
\put(1.2,0){\circle*{0.15}}
\put(-0.075,0.3){}
\put(0.525,0.3){$\da$}
\put(1.,0.3){$\ud$}
\end{picture}

\noindent
\hspace{4.5mm}$\psi_7$\hspace{26.5mm}$\psi_8$\hspace{26.5mm}$\psi_9$\hspace{26.mm}$\psi_{10}$\hspace{25mm}$\psi_{11}$\hspace{25mm}$\psi_{12}$

\vspace{1mm}

\noindent\unitlength1cm
\begin{picture}(1,1)
\put(0,0){\circle*{0.15}}
\put(0.6,0){\circle*{0.15}}
\put(1.2,0){\circle*{0.15}}
\put(-0.15,0.3){$\ud$}
\put(0.525,0.3){$\ua$}
\put(1.125,0.3){}
\end{picture}
\hspace{18mm}
\begin{picture}(1,1)
\put(0,0){\circle*{0.15}}
\put(0.6,0){\circle*{0.15}}
\put(1.2,0){\circle*{0.15}}
\put(-0.15,0.3){$\ud$}
\put(0.525,0.3){}
\put(1.125,0.3){$\ua$}
\end{picture}
\hspace{18mm}
\begin{picture}(1,1)
\put(0,0){\circle*{0.15}}
\put(0.6,0){\circle*{0.15}}
\put(1.2,0){\circle*{0.15}}
\put(-0.075,0.3){$\ua$}
\put(0.45,0.3){$\ud$}
\put(1.125,0.3){}
\end{picture}
\hspace{18mm}
\begin{picture}(1,1)
\put(0,0){\circle*{0.15}}
\put(0.6,0){\circle*{0.15}}
\put(1.2,0){\circle*{0.15}}
\put(-0.075,0.3){}
\put(0.45,0.3){$\ud$}
\put(1.125,0.3){$\ua$}
\end{picture}
\hspace{18mm}
\begin{picture}(1,1)
\put(0,0){\circle*{0.15}}
\put(0.6,0){\circle*{0.15}}
\put(1.2,0){\circle*{0.15}}
\put(-0.075,0.3){$\ua$}
\put(0.525,0.3){}
\put(1.,0.3){$\ud$}
\end{picture}
\hspace{18mm}
\begin{picture}(1,1)
\put(0,0){\circle*{0.15}}
\put(0.6,0){\circle*{0.15}}
\put(1.2,0){\circle*{0.15}}
\put(-0.075,0.3){}
\put(0.525,0.3){$\ua$}
\put(1.,0.3){$\ud$}
\end{picture}

\noindent
\hspace{4.5mm}$\psi_{13}$\hspace{25mm}$\psi_{14}$\hspace{25mm}$\psi_{15}$\hspace{25mm}$\psi_{16}$\hspace{25mm}$\psi_{17}$\hspace{25mm}$\psi_{18}$

\vspace{1mm}

\noindent\unitlength1cm
\begin{picture}(1,1)
\put(0,0){\circle*{0.15}}
\put(0.6,0){\circle*{0.15}}
\put(1.2,0){\circle*{0.15}}
\put(-0.075,0.3){$\da$}
\put(0.525,0.3){$\da$}
\put(1.125,0.3){$\da$}
\end{picture}
\hspace{18mm}
\begin{picture}(1,1)
\put(0,0){\circle*{0.15}}
\put(0.6,0){\circle*{0.15}}
\put(1.2,0){\circle*{0.15}}
\put(-0.075,0.3){$\ua$}
\put(0.525,0.3){$\ua$}
\put(1.125,0.3){$\ua$}
\end{picture}

\noindent
\hspace{4.5mm}$\psi_{19}$\hspace{25mm}$\psi_{20}$

\caption{Schematic representation of the basis functions for three electrons on a three-site lattice.}
\label{fig:basis}
\end{figure*}
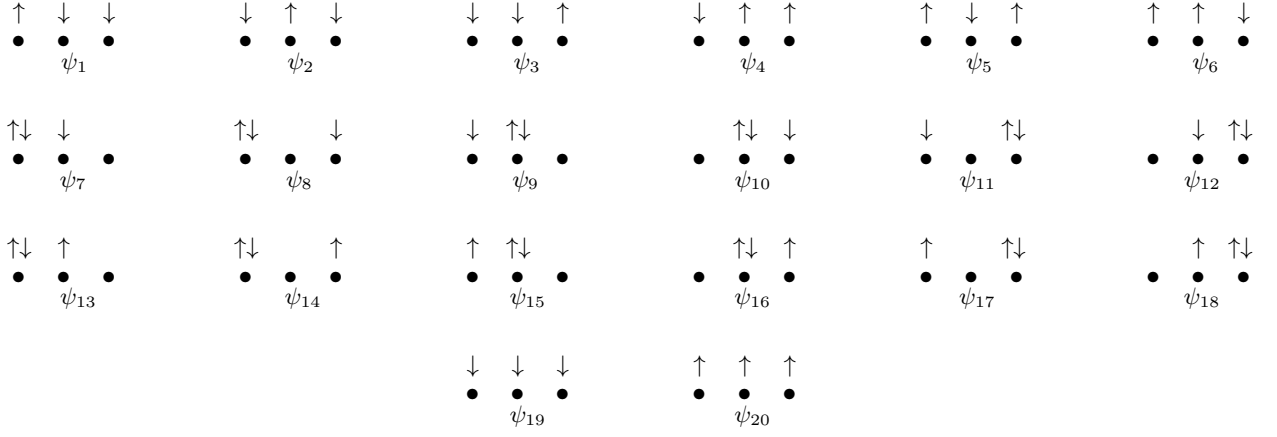

For a three-point lattice, our basis consists of 20 Slater determinants, as schematically illustrated in Fig. \ref{fig:basis}.
Each Slater determinant is an eigenfunction of $\hat S_z$ (spin along the $z$-axis), but not necessarily of $\hat {\bf S}^2$ (total spin): the first 18 Slater determinants,
$\psi_1$ - $\psi_{18}$, have $S_z = \pm 1/2$, and the last two determinants, $\psi_{19}$ and $\psi_{20}$, have $S_z = \pm 3/2$.
By suitable linear combination of the $\psi_n$, one can form 4 quartet wave functions (with $S=3/2$) and 16 doublet wave functions (with $S=1/2$).
These could be used as an alternative basis to diagonalize $\hat H$, but the derivation of the associated Hamiltonian matrix would be more cumbersome.

The space parts of the Slater determinants are constructed using the single-particle basis functions
\begin{equation}\label{varphi}
\varphi_1 = \left( \begin{array}{c} 1 \\ 0 \\ 0\end{array} \right) \;, \hspace{.5cm}
\varphi_2 = \left( \begin{array}{c} 0 \\ 1 \\ 0\end{array} \right) \;, \hspace{.5cm}
\varphi_3 = \left( \begin{array}{c} 0 \\ 0 \\ 1\end{array} \right) \;.
\end{equation}
The column vector notation refers to the three-site lattice, i.e.,
the components of $\varphi_{1,2,3}$ denote the value of
the wave functions on lattice sites 1,2,3. The Slater determinants can then be explicitly written down; for instance,
\begin{equation}
\psi_1(1,2,3) =  \left| \begin{array}{ccc}
\varphi_1(1)\alpha(1) & \varphi_1(2)\alpha(2) & \varphi_1(3)\alpha(3) \\
\varphi_2(1)\beta(1) & \varphi_2(2)\beta(2) & \varphi_2(3)\beta(3) \\
\varphi_3(1)\beta(1) & \varphi_3(2)\beta(2) & \varphi_3(3)\beta(3)
\end{array}\right|,
\end{equation}
and similar for all other $\psi_n$, where $\alpha$ and $\beta$ are up- and down-spinors.

In this notation, the scalar potential operator acting on the $i$th electron is given by
\begin{equation}
V(i)
= \left( \begin{array}{ccc} V_1 & 0 & 0\\0 &V_2 & 0 \\ 0 & 0 & V_3 \end{array} \right)_i \sigma_0(i)\;,
\end{equation}
where the subscript $i$ means that the $3\times 3$ matrix acts on $\varphi_{1,2,3}(i)$, and
$\sigma_0$ is the $2\times 2$ unit matrix (acting on the $\alpha$ and $\beta$ spinors of the $i$th electron).
Here, $\hat V = V(1) + V(2) + V(3)$, and we define the other elements of the three-electron Hamiltonian
$\hat H = \hat T + \hat V + \hat \bfB\cdot {\bm \sigma} + \hat W$ in an analogous manner.

It is easy to see that only the diagonal matrix elements of the scalar potential operator are nonzero, and we have
\begin{eqnarray*}
\ME V_{1,1} = \ldots = \ME V_{6,6} &=& V_1 + V_2 + V_3\\
\ME V_{7,7} = \ME V_{13,13} &=& 2V_1 + V_2 \\
\ME V_{8,8} = \ME V_{14,14} &=& 2V_1 + V_3 \\
\ME V_{9,9} = \ME V_{15,15} &=& V_1 + 2V_2 \\
\ME V_{10,10} = \ME V_{16,16} &=& 2V_2 + V_3 \\
\ME V_{11,11} = \ME V_{17,17} &=& V_1 + 2V_3 \\
\ME V_{12,12} = \ME V_{18,18} &=& V_2 + 2V_3 \\
\ME V_{19,19} = \ME V_{20,20} &=& V_1 + V_2 + V_3 \:.
\end{eqnarray*}

The kinetic energy operator depends on the boundary conditions. In general, we can write
\begin{equation}
T(i)
= \left( \begin{array}{ccc} 0 & -t & -t_p\\-t &0 & t \\ -t_p & -t & 0 \end{array} \right)_i\sigma_0(i),
\end{equation}
where $t_p=0$ for the linear chain and $t_p=t$ for the triangle with periodic boundary conditions.

The only nonzero kinetic energy matrix elements are those that correspond to a single hopping event, without spin flip, namely:
\begin{eqnarray*}
&& \ME T_{1,7} = \ME T_{1,12} = t_p \qquad \ME T_{1,8} = \ME T_{1,10} = t \\
&& \ME T_{2,8} = \ME T_{2,10} = \ME T_{2,11} = -t \qquad \ME T_{2,9}   = t \\
&& \ME T_{3,7} = \ME T_{3,12} = -t_p \qquad \ME T_{3,9} = -t \qquad \ME T_{3,11} = t \\
&& \ME T_{4,13} = \ME T_{4,18} = -t_p \qquad \ME T_{4,14} = \ME T_{4,16} = -t \\
&& \ME T_{5,14} = \ME T_{5,16} = \ME T_{5,17} = t \qquad  \ME T_{5,15}  = -t \\
&& \ME T_{6,13} = \ME T_{6,18} = t_p \qquad \ME T_{6,15} = t \qquad \ME T_{6,17} = -t  \\
&& \ME T_{7,8}  = t \qquad \ME T_{7,9}  = -t \qquad \ME T_{8,11}  = -t_p \quad \\
&& \ME T_{9,10}  = -t_p \qquad \ME T_{10,12}  = \ME T_{11,12}  = -t \\
&& \ME T_{13,14} = t \qquad \ME T_{13,15} = -t \qquad \ME T_{14,17} =  -t_p \\
&& \ME T_{15,16} = -t_p \qquad \ME T_{16,18}  = \ME T_{17,18}  = -t \:.
\end{eqnarray*}
The matrix elements in the lower triangle follow from the hermiticity of the Hamiltonian matrix, i.e.,
$\ME T_{m,n} = \ME T_{n,m}$.

The only nonvanishing matrix elements of the on-site interaction are
the diagonal elements
\begin{displaymath}
\ME W_{7,7}  = \ldots = \ME W_{18,18} =U \:.
\end{displaymath}

The Zeeman terms in the Hamiltonian are given by
\begin{equation}
\bfB(i)\cdot {\bm \sigma}(i) = \left( \begin{array}{ccc} \bfB_1 & 0 & 0\\0 & \bfB_{2}
 & 0 \\ 0 & 0 & \bfB_{3}(i) \end{array} \right)_i\cdot {\bm \sigma}(i) \:.
\end{equation}
We define $\hat B_x = B_x\sigma_x$ and similar for $y$ and $z$.

For the $z$-components,  only the diagonal matrix elements are nonzero, and we get immediately
\begin{eqnarray*}
\MBZ_{1,1}  &=& B_{z1} - B_{z2} - B_{z3}\\
\MBZ_{2,2} &=& -B_{z1} + B_{z2} - B_{z3}\\
\MBZ_{3,3} &=& -B_{z1} - B_{z2} + B_{z3}\\
\MBZ_{4,4} &=& -B_{z1} + B_{z2} + B_{z3}\\
\MBZ_{5,5} &=& B_{z1} - B_{z2} + B_{z3}\\
\MBZ_{6,6} &=& B_{z1} + B_{z2} - B_{z3}\\
\MBZ_{7,7} &=& \MBZ_{12,12} = - B_{z2} \\
\MBZ_{8,8} &=& \MBZ_{10,10} = - B_{z3} \\
\MBZ_{9,9} &=& \MBZ_{11,11} = - B_{z1} \\
\MBZ_{13,13} &=& \MBZ_{18,18} =  B_{z2} \\
\MBZ_{14,14} &=& \MBZ_{16,16} =  B_{z3} \\
\MBZ_{15,15} &=& \MBZ_{17,17} =  B_{z1} \\
\MBZ_{19,19} &=& -B_{z1} - B_{z2} - B_{z3}\\
\MBZ_{20,20} &=& B_{z1} + B_{z2} + B_{z3}
\end{eqnarray*}
For the $x$ and $y$-components, we collect all the spin flips. Defining $B_{\pm j} = B_{xj}\pm i B_{yj}$ for the $j$th lattice site, we get
\begin{eqnarray*}
\MBXY_{1,5} = B_{+3}  &\qquad&  \MBXY_{1,6} =  B_{+2}  \\
\MBXY_{1,19} = B_{-1}  &\qquad& \MBXY_{2,4}  = B_{+3}  \\
\MBXY_{2,6} = B_{+1}  &\qquad& \MBXY_{2,19} = B_{-2} \\
\MBXY_{3,4}  =  B_{+2} &\qquad& \MBXY_{3,5}  = B_{+1} \\
\MBXY_{3,19} = B_{-3} &\qquad& \MBXY_{4,20}  = B_{+1} \\
\MBXY_{5,20}  = B_{+2}  &\qquad& \MBXY_{6,20}  = B_{+3}  \\
\MBXY_{7,13}  = B_{+2}  &\qquad& \MBXY_{8,14} = B_{+3}  \\
\MBXY_{9,15}  = B_{+1} &\qquad& \MBXY_{10,16} = B_{+3} \\
\MBXY_{11,17}  = B_{+1} &\qquad& \MBXY_{12,18}  = B_{+2}
\end{eqnarray*}
The matrix elements in the lower triangle follow directly as $\MBXY_{m,n} = \MBXY_{n,m}^*$.

From the eigenvectors of Eq. (\ref{M}), we obtain the spin-density matrix from the following formula:
\begin{equation}\label{sdm}
\underline{\underline n}(1) = \sum_{n,m=1}^{20} c_n c_m^* \! \sum_{2,3 \atop(space)} \mbox{tr}_{2,3 (spin)}[ \psi_n(1,2,3) \psi_m^\dagger(1,2,3)].
\end{equation}
The final result is
\begin{eqnarray*}
n_{\uu} &=& |c_1|^2 \varphi_1 + |c_2|^2 \varphi_2 + |c_3|^2 \varphi_3 + |c_4|^2 (\varphi_2+\varphi_3) \\
&+& |c_5|^2 (\varphi_1+\varphi_3) + |c_6|^2 (\varphi_1+\varphi_2) + |c_7|^2 \varphi_1 \\
&+& |c_8|^2 \varphi_1 + |c_9|^2 \varphi_2 + |c_{10}|^2 \varphi_2 + |c_{11}|^2 \varphi_3 \\
&+& |c_{12}|^2 \varphi_3 +
|c_{13}|^2 (\varphi_1+\varphi_2) + |c_{14}|^2 (\varphi_1+\varphi_3)  \\
&+&
|c_{15}|^2 (\varphi_1+\varphi_2)
+ |c_{16}|^2 (\varphi_2+\varphi_3) + |c_{17}|^2 \!(\varphi_1+\varphi_3)\\
&+&  |c_{18}|^2 (\varphi_2+\varphi_3)
+ |c_{20}|^2 (\varphi_1 + \varphi_2 + \varphi_3)
\end{eqnarray*}
\begin{eqnarray*}
n_{\dd} &=& |c_1|^2 (\varphi_2+\varphi_3) + |c_2|^2 (\varphi_1+\varphi_3) + |c_3|^2 (\varphi_1+\varphi_2)\\
& +& |c_4|^2 \varphi_1 + |c_5|^2 \varphi_2 + |c_6|^2 \varphi_3
+ |c_7|^2 (\varphi_1+\varphi_2) \\
&+&
|c_8|^2 (\varphi_1+\varphi_3) + |c_9|^2 (\varphi_1+\varphi_2) + |c_{10}|^2 (\varphi_2+\varphi_3)\\
& +& |c_{11}|^2(\varphi_1+ \varphi_3) + |c_{12}|^2 (\varphi_2+\varphi_3) \\
&+&
|c_{13}|^2 \varphi_1 + |c_{14}|^2 \varphi_1 +|c_{15}|^2 \varphi_2
+ |c_{16}|^2 \varphi_2\\
& +& |c_{17}|^2\varphi_3 + |c_{18}|^2\varphi_3
+ |c_{19}|^2 (\varphi_1 + \varphi_2 + \varphi_3)
\end{eqnarray*}
\begin{eqnarray*}
n_{\ud} &=& c_1c_{19}^* \varphi_1 + c_2 c_{19}^* \varphi_2 + c_3 c_{19}^* \varphi_3
+ c_4 c_2^* \varphi_3 + c_4 c_3^* \varphi_2\\
& +& c_5 c_3^* \varphi_1
+ c_5 c_1^*\varphi_3 + c_6 c_1^*\varphi_2 +c_6 c_2^*\varphi_1 + c_{13}c_7^* \varphi_2 \\
&+& c_{14}c_8^* \varphi_3 + c_{15}c_9^* \varphi_1 + c_{16}c_{10}^* \varphi_3 + c_{17}c_{11}^* \varphi_1\\
& +& c_{18}c_{12}^* \varphi_2 +c_{20}c_4^* \varphi_1 + c_{20}c_5^* \varphi_2 +c_{20}c_6^* \varphi_3
\\[3mm] 
n_{\du} &=& n_{\ud}^* \:.
\end{eqnarray*}
From this, one obtains the density as $n = n_{\uu} + n_{\ud}$
and the $x-$, $y-$ and $z$-components of the magnetization as
$m_x = n_{\ud} + n_{\du}$, $m_y = i(n_{\ud} - n_{\du})$, and $m_z = n_{\uu} - n_{\dd}$.

\bibliography{Trimer_refs}

\begin{thebibliography}{36}
\bibitem [{foo()}]{footnote}%
  \BibitemOpen
  \href@noop {} {}\bibinfo {note} {The Hubbard dimer with three electrons seems
  an even simpler choice, but due to particle-hole symmetry it is equivalent to
  a Hubbard dimer with a single electron.}\BibitemShut {Stop}%
\end{thebibliography}

\end{document}